# Bayesian Co-Navigation of a Computational Physical Model and AFM Experiment to Autonomously Survey a Combinatorial Materials Library

Boris N. Slautin,[1,*] Kamyar Barakati,[1] Yu Liu,[1] Reece Emery[1],

Philip Rack,[1] and Sergei V. Kalinin[1,2*]

[1] Department of Materials Science and Engineering, University of Tennessee, Knoxville, TN 37923, USA

[2] Pacific Northwest National Laboratory, Richland, WA 99354, USA

Building autonomous experiment workflows requires transcending beyond the data-driven surrogate models to incorporate and dynamically refine physical theory during exploration. Here we demonstrate the first fully automated experimental realization of Bayesian co-navigation – a framework in which an autonomous agent simultaneously runs a physical experiment and a computationally expensive physical model. Using an automated AFM platform coupled to a kinetic Monte Carlo (kMC) model of thin-film growth, the system infers a set of effective bond energies for the $(CrTaWV)_x$–$Mo_{(1-x)}$ pseudo-binary combinatorial library, progressively adjusting the kMC parameters to decrease the epistemic disparity between simulation and experiment. This real-time theoretical refinement enables the kMC model to capture the behavior of the specific materials system and reveals the mechanistic role of hetero-bonding in governing surface diffusion. Together, these results establish co-navigation as a general strategy for tightly integrating physical models with autonomous experimental platforms to produce interpretable and continually self-correcting theoretical modelling of complex materials systems.

[*] Authors to whom correspondence should be addressed: bslauti1@utk.edu and sergei2@utk.edu

Recent years have seen rapidly growing interest in active and automated experiments across materials synthesis, characterization, scattering, microscopy, and related fields, driven by the convergence of high-throughput instrumentation, advanced control hardware, and machine learning.[1-17] Autonomous workflows are increasingly explored as a means to overcome the limitations of traditional, manually driven experimentation, where human intuition and serial trial-and-error often constrain the rate and breadth of discovery. In microscopy, active approaches promise to optimize imaging parameters, adaptively select regions of interest, and uncover rare or transient phenomena,[9,18-23] while in materials synthesis they enable closed-loop exploration of large compositional and processing spaces.[4,17,24-27] Together, these developments motivate a new class of experimental paradigms in which algorithms and instruments jointly navigate complex design spaces, transforming how we generate, interpret, act, and learn from experimental data.

State-of-the-art autonomous experimentation in materials and microscopy is largely built around Bayesian optimization (BO).[28-31] In BO, an unknown objective function such as a figure of merit for material performance or image quality is approximated by a probabilistic surrogate model, which is iteratively updated as new measurements are acquired. The surrogate is typically either purely data-driven, for example a zero-mean Gaussian process (GP) that learns the response surface solely from observed data,[2,22,23,32-37] or it can incorporate prior physical knowledge through a probabilistic model encoded as a symbolic expression, as in structured Gaussian processes.[38] In both cases, an acquisition function uses the surrogate's mean and uncertainty to select the next experiment, balancing exploration of poorly known regions with exploitation of promising candidates, thereby enabling sample-efficient optimization of expensive experiments.

The BO-based approaches are widely encoded in mainstream open-source libraries such as BOTORch,[39] Ax,[40] Baybe[41] etc. facilitating their widespread adoption. Recent works have begun to integrate large language models (LLMs) into the Bayesian Optimization (BO) loop. Among other examples, LLMs can assist GP-based BO by refining acquisition functions,[42,43] constructing domain-aware priors,[44,45] incorporating real-time human feedback into the optimization process.[46] Some recent methods suggest completely substituting the GP surrogate with an LLM, although such an approach relies on heuristic uncertainty estimation rather than a principled probabilistic model.[47-49] In materials and chemical science, LLM-augmented BO has been applied to catalyst optimization,[49] reaction yield tuning,[50] lithium-ion battery parameter identification,[51] and molecular/material discovery.[52]

Actively incorporating domain knowledge, whether through probabilistic models or LLM-based reasoning, can substantially accelerate optimization and enhance the effectiveness of autonomous experimentation. In principle, the most powerful form of such integration would rely on physics-based models, which not only offer a cost-efficient pathway for materials design but also enable the discovery of underlying physical mechanisms that govern material behavior. At the same time, a very common scenario arises in which the theoretical model is available only in numerical form and is itself computationally expensive to evaluate. One possible solution lies in the construction of the fast low-fidelity surrogate models capable of replicating theoretical model behavior within the parameter space as mean function of Gaussian Process.[53] However, this approach brings the challenge of epistemic uncertainty of a fast low-fidelity surrogate. Correspondingly, of interest is the development of workflows that allow one to orchestrate expensive experiment and high-fidelity computationally expensive calculations to replicate specific material behavior over common exploration space.

We note that this problem setting is very different from multifidelity[54-56] and multitask BO.[57] In multifidelity BO, cheaper, lower-fidelity models are used as approximations to a single high-fidelity objective, with the primary challenge being how to allocate queries across fidelities to accelerate optimization of that objective. In multitask BO, a shared latent surrogate is constructed for several related tasks (e.g., different materials, operating conditions, or target properties), exploiting correlations to improve sample efficiency across tasks. By contrast, in our co-navigation setting theory and experiment represent distinct but coupled domains, both expensive, with the central aim of jointly inferring model parameters and steering experimental exploration rather than merely choosing among information sources or transferring knowledge across tasks, as recently illustrated for the pre-acquired Piezoresponse Force Microscopy data[58] with FerroSim lattice model[59] as a theoretical model.

Here we demonstrate the fully automated co-navigation experiment for exploring surface-topography evolution in a $(CrTaWV)_x$-$Mo_{(1-x)}$ pseudo-binary combinatorial library. In this workflow, a kinetic Monte Carlo (kMC) model of surface evolution serves as the theoretical component and progressively evolves into a material-specific "digital twin" by continuously inferring and updating its parameters from experimental observations. The experimental surface roughness measurements were performed by automated atomic force microscopy (AFM).[22] The resulting adjusted model of material behavior realized as the physical model of composition-dependent phase and surface roughness evolution, enables downstream modeling of films with targeted surface characteristics.

**I. Principle of the co-navigation experiment**

Bayesian co-navigation inherits key ideas from the Kennedy & O'Hagan calibration framework[60] and projects them into an active-learning paradigm suitable for autonomous experimentation. This is a multi-loop active-learning framework designed to align theoretical modeling and experimental exploration in real time.[58] Instead of treating theory as a fixed prior and experiment as a static source of data, co-navigation orchestrates three concurrent Bayesian optimization cycles: a theoretical loop, an experimental loop, and an outer theory-update loop that links them (Figure 1). The experimental and theoretical loops each build a Gaussian-process or Deep Kernel Learning surrogate model over their respective object spaces. These surrogates guide the selection of new experiments or simulations through uncertainty-based acquisition functions, enabling rapid and targeted exploration the objective spaces. The use of surrogates allows to avoid the need for theory and experiment be sampled in the same regions of the parameter spaces, allows balancing the throughputs of the cycles, nature of updates (batch vs. single point), and the use of individual predictive uncertainties to guide the exploration.

Crucially, the mapping from the experimental object space to the measured observables must be defined in a one-to-one and comparable manner for both experiment and theory: for every point explored experimentally, the theoretical model must generate a prediction of the same physical quantity, such as roughness, intensity, yield, or a full spectral response. In our case, the surface roughness represents the target functionality discovered in compositional space of combinatorial library.

The key component is the outer theory-update loop (T-loop). At regular intervals, it evaluates the mismatch between the experimental surrogate and the theoretical surrogate by comparing predicted target properties. This mismatch, expressed as a mean-squared error over corresponding experimental and theoretical objects, defines the epistemic uncertainty between theory and experiment. A Bayesian optimization routine then explores the hyperparameter space of the theoretical model to find values that minimize this discrepancy. Each update redefines the theoretical object space, the simulation outputs, and consequently the surrogate model on which the T-loop operates.

Through this iterative interaction, Bayesian co-navigation achieves dynamic calibration of a theoretical model during an ongoing autonomous experiment. The algorithm systematically reduces the mismatch between predicted and measured functionality, progressively refining the general theoretical model into an accurate, system-specific representation. Unlike classical Bayesian calibration frameworks, which operate offline with fixed datasets,[60] co-navigation performs model updating, experimental design, and theoretical exploration continuously and adaptively.[58] This enables theory and experiment to co-evolve, creating a closed-loop discovery workflow capable of steering experiments, refining models, and uncovering mechanism-level insights with minimal human intervention.

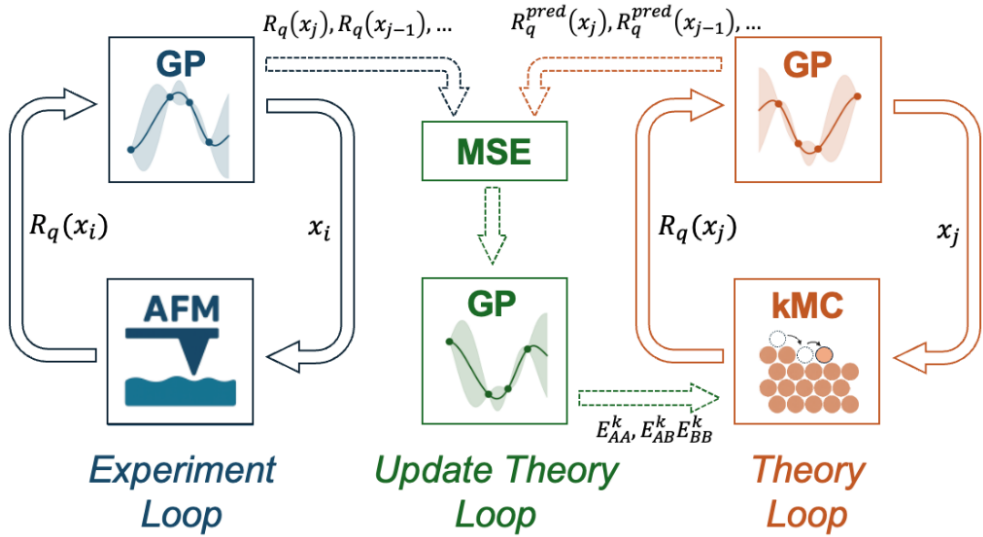

**Figure 1.** A schematic overview of the co-navigation framework, shown for the case of autonomous AFM measurements coupled to a kinetic Monte Carlo model of thin-film growth. The agent coordinates experimental exploration, theoretical simulations, and outer-loop model calibration, enabling dynamic reduction of epistemic discrepancies between experiment and theory, thereby constructing a theoretical representation of the specific material system.

**II. Models and ground truth behavior**

As a first step, we outline the theoretical models and the ground-truth material behavior that serve as baselines for the co-navigation experiment. The overall goal of this experiment was to jointly navigate the theoretical model and the experimental measurements to investigate the compositional evolution of surface roughness in a spreading combinatorial library.

**II.a. Material system.** As a model system, we employed a $(CrTaWV)_x$–$Mo_{(1-x)}$ spread combinatorial library with composition varying continuously from ~0.15 to 0.85 across the film. The psuedo-binary $(CrTaWV)_x$–$Mo_{(1-x)}$ combinatorial library was synthesized via co-sputtering an alloyed target of roughly equiatomic CrTaWV target and a pure Mo target. The system was pumped to ~ $5 \times 10^{-7}$ Torr and backfilled with 25 sccm Ar and throttled to a sputtering pressure of 5 mTorr. To achieve roughly equal deposition rates, the CrTaWV target was powered to 200 W and the Mo target was 120W. The growth temperature was 500 °C and a total time of 3 hours and 20 minutes for a total thickness of ~2 μm. Although deviations from perfect linearity in the compositional profile are possible, we treat the library as having a nominally linear composition gradient along the primary axis for the purposes of this study.

To establish the ground-truth behavior of surface roughness, we first examined its evolution across the compositional gradient. Mean-square surface roughness was used as the primary metric and was measured by AFM operated in non-contact mode (Figure 2). For ground-truth acquisition, we performed 100 evenly spaced non-contact topography scans spanning the full compositional range. Three scans were identified as outliers and excluded, while the remaining measurements provided a reliable dataset for reconstructing the roughness profile. Notably, the full automated acquisition required more than 16 hours, underscoring both the impracticality of exhaustive grid-based measurements for downstream characterization and the need for faster, physics-aware methods to enable such studies.

At the molybdenum-rich end of the compositional space (x = 0.15), the library exhibits a well-defined granular microstructure with an average surface roughness of approximately 20–23 nm. As composition varies, both the characteristic grain size and the mean roughness gradually increase, reaching a maximum near x ≈ 0.7. Beyond this threshold, in the CrTaWV-rich (high-x) region, the surface roughness decreases sharply, accompanied by the disappearance of the granular morphology (Figure 2).

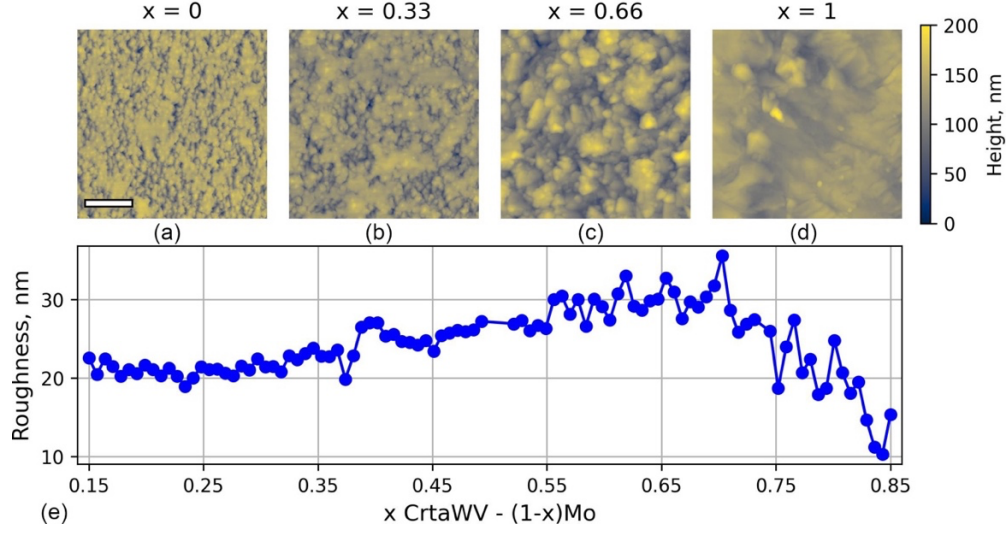

. (a–d) AFM topography scans measured at different compositions. (e) Compositional profile of the RMS roughness across the full compositional space. Scale bar: 1 μm.

**II.b. Theoretical model.** A kinetic Monte Carlo (kMC) model is employed to provide a theoretical description of the evolution of surface roughness across the compositional gradient.[61,62] In this lattice-based formulation, thin film growth is modeled as a sequence of thermally activated diffusion events in which adatoms hop between neighboring lattice sites with physically motivated rates. Each possible hop is assigned an Arrhenius-type rate

$$r_{i \to j} = \nu_0 \exp\left(-\frac{E_{diff}(i \to j)}{k_B T}\right), \quad (1)$$

where $\nu_0$ is the attempt frequency, $T$ is the substrate temperature, and $E_{\text{diff}}$ is the activation barrier for the given hop. The kMC algorithm statistically selects events according to these rates, thereby capturing the stochastic nature of adatom motion during film deposition.

The diffusion barrier $E_{\text{diff}}$ is constructed using a broken-bond model that accounts for the local bonding environment of the diffusing atom. During a hop, the atom breaks a set of A-A, B-B, and A-B bonds with associated energetic penalties, leading to

$$E_{diff} = E_0 + n_{AA}E_{AA} + n_{BB}E_{BB} + n_{AB}E_{AB} + E_{step}, \quad (2)$$

where $E_0$ is the intrinsic terrace diffusion barrier, $E_{XY}$ are effective bond energies, and $n_{XY}$ denote the number of bonds broken during the hop. These terms govern alloy mixing, local ordering, and the stability of surface clusters as a function of composition. The model also incorporates step-edge effects: hops down a step are energetically favored (slip-down), while hops up a step encounter an Ehrlich–Schwoebel barrier. Thus,

$$E_{step} = \begin{cases} -\Delta E_d, & \Delta h < 0 \\ 0, & \Delta h = 0 \\ \gamma_{ES}\Delta h, & \Delta h > 0 \end{cases} \tag{3}$$

where $\Delta h$ is the local height difference and $\gamma_{ES}$ is a coefficient defining an Ehrlich–Schwoebel barrier proportional to the heights difference. These contributions regulate the balance between mound formation, terrace smoothing, and layer-by-layer growth.

Together, these elements yield a physically grounded kMC description of thin-film deposition that captures key microscopic processes, including diffusion, step-edge kinetics, and alloy-dependent bonding energies, that collectively determine the emergent surface roughness. This model provides the theoretical counterpart to the AFM characterization and enables a direct, composition-resolved comparison between simulation and experiment within the co-navigation framework.

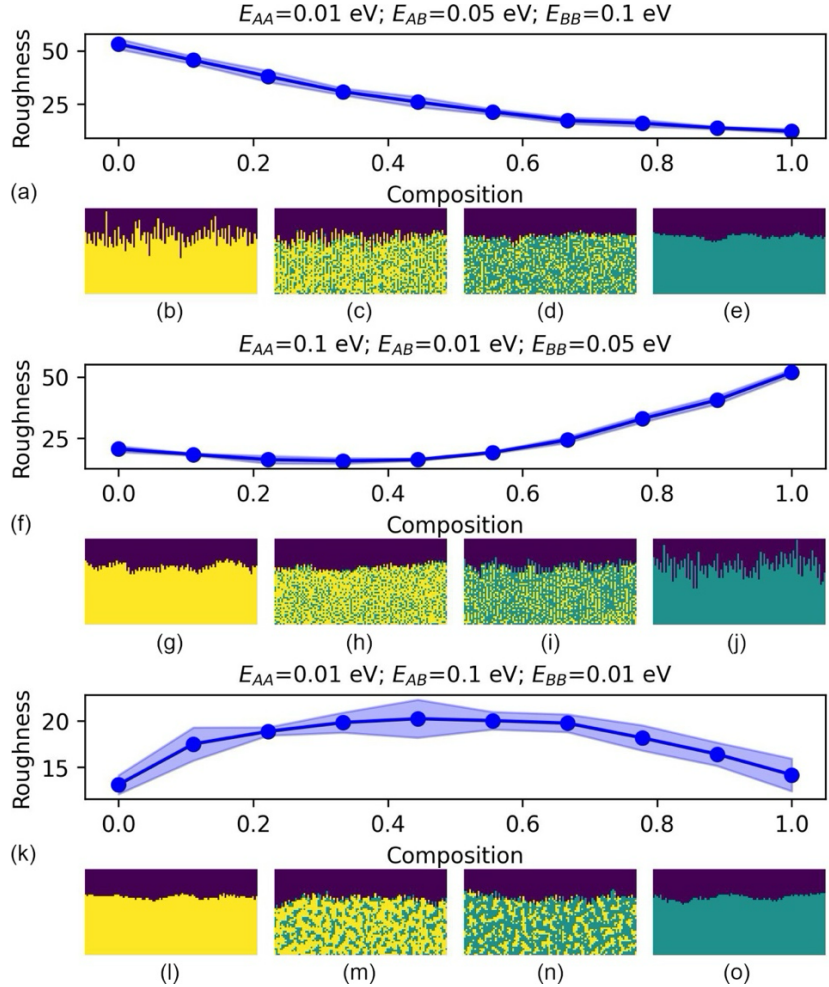

**Figure 3.** Compositional dependencies of surface roughness in kMC-simulated films for different bond energies. Panels (a), (f), and (k) show roughness as a function of composition for different

hyperparameter sets. Panels (b–e), (g–j), and (l–o) present representative microstructures at (b, g, l) x = 0, (c, h, m) x = 0.33, (d, i, n) x = 0.66, and (e, j, o) x = 1.

### III. Realization of the co-navigation approach

To implement the theoretical component of the co-navigation framework, we employ a two-dimensional lattice-based Kinetic Monte Carlo (kMC) model that captures the essential physics of thin-film sputtering and composition-dependent surface evolution. The film surface is represented as a 2D lattice with a lateral width of 300 sites to balance computational complexity with the ability to reproduce mesoscale morphological features.

Sputtering in the model is represented as a constant flux of incoming particles. A fully detailed kMC description of the (CrTaWV)x–Mo(1-x) system would require tracking all constituent atomic species and their interaction parameters. However, the quasi-binary structure of the experimental library makes these parameters strongly interdependent and non-identifiable. To ensure that the model complexity remains aligned with the information content of the experiments, we adopt a simplified binary representation with two effective species, A and B, corresponding to the compositional extremes of the combinatorial gradient. Component A represents the effective composition on the CrTaWV-rich side of the library ($x = 0.85$), and component B represents the Mo-rich side ($x = 0.15$). Therefore, experimental compositions are modeled as an $A_x B_{1-x}$ mixture, where kMC particles correspond to effective structural units rather than individual atoms. In the experimental section below, we use a renormalized composition scale in which $x = 0$ corresponds to the $(CrTaWV)_{0.85}$–$Mo_{0.15}$ end of the library, and $x = 1$ corresponds to the $(CrTaWV)_{0.15}$–$Mo_{0.85}$ end.

At a given composition $x$, the incoming flux is sampled according to the species probability distribution, ensuring that the simulated film directly reflects the experimental composition. A total of $10^4$ atoms were deposited in each kMC simulation. To enable direct comparison with experimental data, the simulated roughness values were multiplied by an empirical scaling factor of 5, bringing the model output into the same numerical range as the AFM-measured roughness in nanometers. The following fixed parameters were used in the kMC model: $\Delta E_d = 0.1$ eV, $\gamma_{ES} = 0.15$ eV, and $E_0 = 0.2$ eV. The effective bond energies $E_{XY}$ served as the hyperparameters of the theoretical model and were constrained to the range 0.01–0.1 eV. The overarching goal of the autonomous co-navigation experiment was to learn these $E_{XY}$ values so that the kMC model accurately reproduced the experimentally observed roughness evolution across the compositional gradient.

Both the theoretical and experimental loops were initiated with roughness data obtained at five randomly selected compositions, followed by an automated uncertainty-driven exploration cycle. Initial hyperparameter values for the kMC model were chosen randomly. Experimental roughness measurements were performed after every 10 theoretical simulations. Theory hyperparameters were also updated every 10 iterations: the first three updates were sampled randomly, while subsequent updates were guided by the Low Confidence Bound (LCB) acquisition function, which govern the optimization toward reducing the mean squared error (MSE) between experimental roughness observations and theoretical predictions for corresponding compositions. Because updates to the theory hyperparameters make earlier kMC simulations inconsistent with the most recent model state, the theoretical surrogate would become invalid if trained on all past simulations. To address this, we introduced a memory tail mechanism: at each update cycle, the GP surrogate for the theoretical model was constructed using only the most recent 10 simulations. This approach ensures that the surrogate is always consistent with the current set of model hyperparameters and enables stable convergence of the outer theory-optimization loop. The total number of iterations in automated experiment were 201.

AFM measurements were performed using a Jupiter AFM (Oxford Instruments) operated in non-contact topography mode. A hard silicon cantilever Tap300Al-G (BudgetSensors) with a measured spring constant of 32 N/m was used to ensure stable imaging. All scans were acquired under fully automated control using the AESPM Python library, which handled probe positioning, surface approach, scan execution, and data acquisition without human intervention. During co-navigation, the algorithm dynamically selected measurement locations within the 100 mm $(CrTaWV)_x–Mo_{(1-x)}$ combinatorial library based on the co-navigation model prediction. complete co-navigation run required approximately two hours, including all AFM scans, theoretical simulations, and model training and predictions.

The co-navigation code is publicly available in GitHub repository. In addition, we provide there a Google Colab notebook that enables users to run the co-navigation workflow in a fully simulated mode using grid-based measurements for experiment emulation.

**IV. Experiment**

To assess the effectiveness of the co-navigation process, we monitor the evolution of the surrogate model predictions as the experiment progresses (Figure 3). The uncertainty-driven experimental exploration of the roughness across the compositional gradient rapidly converges to the characteristic U-shaped profile with a pronounced maximum near $x \approx 0.7$, in agreement with the ground-truth measurements. Importantly, the validity of each AFM measurement is preserved throughout the run, because experimental data remain fully consistent regardless of changes in the theoretical model. As a result, the co-navigation framework generally requires far fewer experimental iterations than theoretical simulations. In our implementation, the ratio between theoretical simulations and AFM measurements was 10:1.

The large number of simulations during the early stages of the run reflects the need to update the theoretical model; changes in the $E_{XY}$ bond energies render earlier kMC simulations inconsistent with the current model state. This inconsistency diminishes as the theory-update loop converges and disappears once the model hyperparameters stabilize. The simulated roughness values corresponding to outdated hyperparameter settings are shown as gray points (Figure 4). Their wide spread illustrates two important effects: (i) the strong influence of bond energies on the overall shape of the compositional roughness profile, and (ii) the active exploration of the parameter space by the algorithm, which naturally produces diverse roughness–composition dependencies during the early optimization phases.

The overall evolution of the co-navigation experiment results in a convergence of the theoretical model toward the experimentally observed U-shaped roughness profile. The remaining deviations between theoretical predictions and measured values are expected and arise from the simplicity of the chosen kMC model and the assumptions embedded within it. Nevertheless, the model successfully captures the essential features of the roughness evolution, including the global profile shape and the location of the maximum near $x \approx 0.7$. This agreement demonstrates the effectiveness of the co-navigation framework in steering a simplified theoretical model toward accurate representation of experimentally observed behavior. The complete evolution of the theoretical model, along with the progression of both theoretical and experimental exploration during the automated co-navigation run, is provided as an animated GIF in the Supplementary Materials (see Data availability). A video recording of the experimental run is also included.

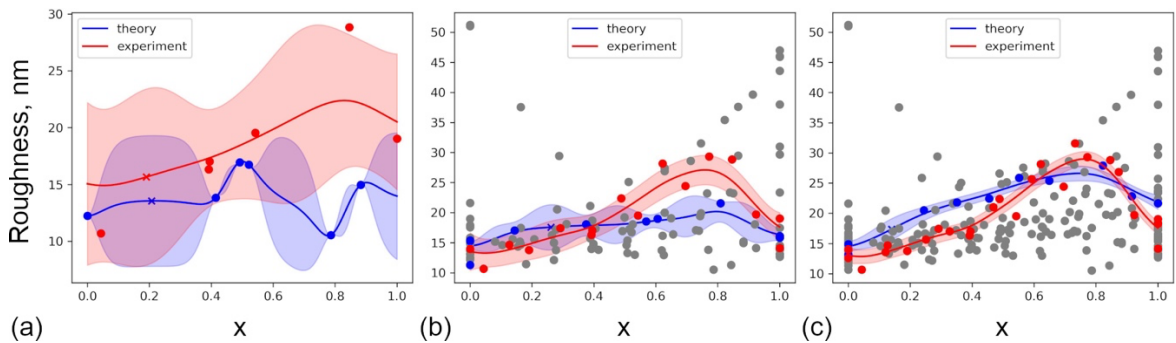

**Figure 4.** Theoretical and experimental roughness estimates across the compositional space, shown together with the corresponding theory and experiment surrogate model predictions at different stages of the automated co-navigation process: (a) iteration 1, (b) iteration 100, and (c) iteration 199.

To analyze the tuning of the KMC model during the theory-update loop, we track the evolution of the MSE and the changes in the bond energies throughout the experiment (Figure 5a). During the initial theory-update iterations, the MSE exhibits significant fluctuations, reflecting the randomized hyperparameter selection and the exploratory behavior driven by the LCB acquisition function. As the optimization proceeds, these fluctuations diminish and the MSE gradually decreases, showing only minor deviations. This transition from an unstable exploratory phase to a stable exploitation regime matches the expected behavior of the co-navigation framework as the model parameters converge toward their optimal values.

The evolution of the hyperparameters themselves provides additional insight into the co-navigation dynamics. Because the theory-update loop operates in a three-dimensional space defined by the bond energies, direct visualization of the corresponding surrogate model predictions becomes impractical. Instead, we analyze the trajectories of the hyperparameters in this energy space (Figure 5b,c), as well as the step-to-step distances (Figure 5d) between sequent hyperparameter sets. The behavior of all three bond energies correlates closely with the MSE evolution and follows the expected co-navigation pattern. During the early stages, the bond energies exhibit large jumps between iterations, reflecting active exploration of the parameter space. As the experiment progresses, these fluctuations decrease, and by approximately iteration 150 the hyperparameters converge toward a consistent pattern in which $E_{AB}$ exceeds both intra-type bond energies. Although adjustments continue throughout the remainder of the run, this overall structure remains stable.

Analysis of the exploration trajectory in the energy space shows that, after the initial broad search, the selected hyperparameter sets become confined to a narrow region (Figure 5c). This behavior is also evident in the monotonic reduction of the distance between consecutive hyperparameter vectors as the experiment proceeds, confirming a transition from exploration to exploitation and the gradual stabilization of the theoretical model (Figure 5d).

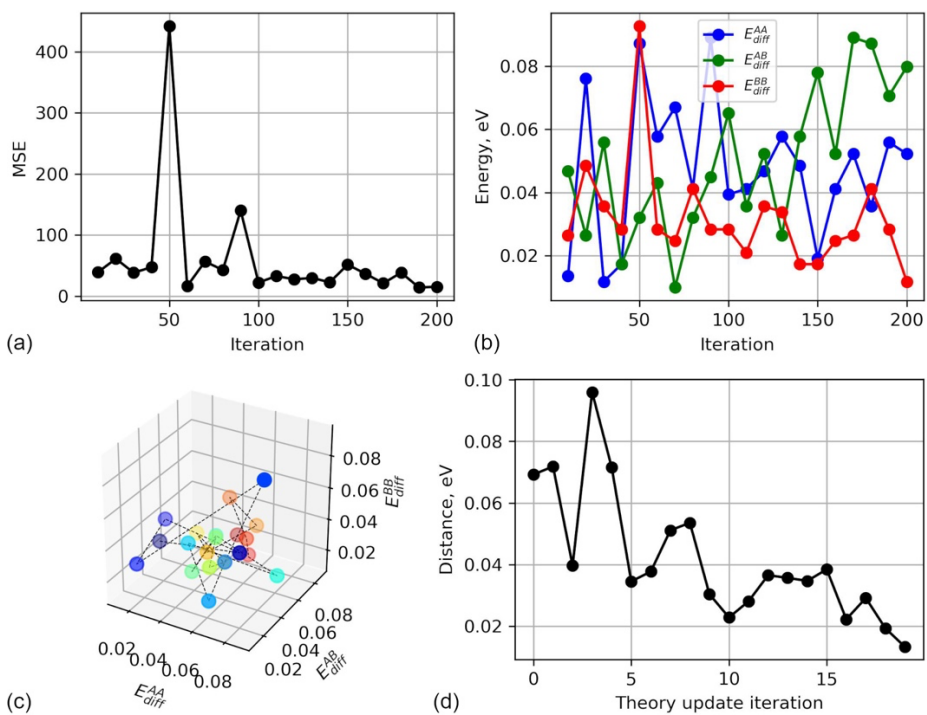

**Figure 5.** Evolution of the theoretical model hyperparameters during the co-navigation experiment. (a) Mean squared error (MSE) between experimental and theoretical roughness. (b) Effective bond energy parameters $E_{AA}$, $E_{BB}$, and $E_{AB}$. (c) Trajectory of the hyperparameter updates in the three-dimensional bond-energy space, with color indicating iteration number (blue → red). (d) Evolution of the distance between sequent hyperparameter sets, illustrating the transition from exploration to convergence.

The optimized hyperparameters follow the pattern $E_{AB} > E_{AA}, E_{BB}$, where A denotes the CrTaWV-rich species and B corresponds to Mo. In our simple lattice model, these can be attributed to *effective* bond energies between sites occupied by CrTaWV-rich (A) and Mo-rich (B) atoms, rather than elemental pair interactions. The fact that $E_{AB}$ exceeds both $E_{AA}$ and $E_{BB}$ implies that hetero-bonds A-B are energetically more stable than either A-A or B-B intra-bonds. At intermediate compositions, the density of A-B bonds is high, so surface diffusion is hindered and the surface demonstrates a pronounced roughness maximum. Toward the Mo-rich end, the surface is dominated by weaker B-B interactions, allowing sufficient diffusion to form a granular low-roughness morphology. At the CrTaWV-rich end, AFM reveals a mostly smooth surface with no grains, consistent with amorphous growth. Together, the bond-energy pattern and the resulting morphology evolution support a picture of CrTaWV–Mo as a system that energetically favors mixed local environments, with maximal roughness emerging where hetero-bonds are abundant and strongest.

**V. Summary**

This work demonstrates the first fully automated experimental realization of Bayesian co-navigation. By coupling autonomous AFM measurements with a kMC model of thin-film growth, we showed how a computationally expensive physical model can be dynamically calibrated to experimental observations in real time. Through this process, the co-navigation framework produced a material specific theoretical representation capable of accurately reproducing the surface-roughness evolution across a combinatorial materials library. Beyond matching the experimental behavior, the resulting model also revealed mechanistic insights into the origin of the observed roughness dependence, specifically the influence of composition-dependent bonding energies on surface diffusion and morphology formation. These insights underscore the value of co-navigation in uncovering physically meaningful relationships that are difficult to extract from experiments or simulations alone and highlight the broader potential of this approach for integrating physical models with autonomous experimentation across a wide range of materials systems.

**Acknowledgements**

This material (BNS, SVK) is based upon work supported by the National Science Foundation under Award No. NSF 2523284. The combinatorial library growth (RE, PDR) was supported by the National Science Foundation through the Materials Research Science and Engineering Center program at the UT Knoxville Center for Advanced Materials and Manufacturing under Award No. DMR-2309083.

**Author contributions**

**BNS:** Conceptualization; Software; Data curation; Writing – original draft. **KB**: Data curation; Writing – review & editing. **YL:** Data curation; Writing – review & editing. **RE:** Resources; Writing – review & editing. **PR:** Resources; Writing – review & editing. **SVK:** Conceptualization; Supervision; Writing – review & editing.

**Data availability**

Raw experimental data and the supporting codes can be accessed through the GitHub repository: https://github.com/Slautin/2025_Co-navigation_CrTaWV-Mo.git